\documentclass[preprintnumbers]{revtex4}

\usepackage{graphicx}
\def\ie{{\it i.e.}}

\def\etal{{\it et al.}}

\def\mpl{\ifmmode \overline M_{Pl}\else $\overline M_{Pl}$\fi}
\def\to{\rightarrow}
\newskip\zatskip \zatskip=0pt plus0pt minus0pt
\def\matth{\mathsurround=0pt}
\def\lsim{\mathrel{\mathpalette\atversim<}}

\def\atversim#1#2{\lower0.7ex\vbox{\baselineskip\zatskip\lineskip\zatskip
  \lineskiplimit 0pt\ialign{$\matth#1\hfil##\hfil$\crcr#2\crcr\sim\crcr}}}

\begin{document}
\bibliographystyle{revtex}

\preprint{SLAC-PUB-9072/
          P3-38}

\title{Radion Mixing Effects on the Properties of the Standard Model 
Higgs Boson}

\author{J.L.Hewett and T.G. Rizzo}

\email[]{hewett,rizzo@slac.stanford.edu}
\affiliation{Stanford Linear Accelerator Center, 
Stanford University, Stanford, California 94309 USA}

\date{\today}

\begin{abstract}
We examine how mixing between the Standard Model(SM) Higgs boson, $h$, and 
the radion 
of the Randall-Sundrum model modifies the expected properties of the Higgs 
boson. In particular we demonstrate that the total and 
partial decay widths of the 
Higgs, as well as the $h\to gg$ branching fraction,
 can be substantially altered from their SM expectations, while
the remaining branching fractions are modified less than $\lsim 5\%$
for most of the parameter region. 
\end{abstract}

%\maketitle must follow title, authors, abstract and \pacs
\maketitle

%\section{Introduction}

The Randall-Sundrum(RS) model{\cite {rs}} offers a potential solution to the 
hierarchy problem that can be tested at present and future 
accelerators{\cite {dhr}}. In this model the SM fields lie on one of two 
branes that are embedded in 5-dimensional AdS space 
described by the metric 
$ds^2=e^{-2ky}\eta_{\mu\nu}dx^\mu dx^\nu-dy^2$, where $k$ is the 5-d curvature 
parameter of order the Planck scale. To solve the hierarchy problem 
the separation between the two branes, $r_c$, must have a value of 
$kr_c \sim 11-12$. That this quantity can be stabilized and made natural has 
been demonstrated by a number 
of authors{\cite {gw}} and leads directly to the existence of a radion($r$), 
which corresponds to a quantum excitation of the brane separation. It can be 
shown that the radion couples to the trace of the 
stress-energy tensor with a strength $\Lambda$ of 
order the TeV scale, \ie, ${\cal L}_{eff}=-r~T^\mu_\mu /\Lambda$. 
(Note that $\Lambda= \sqrt 3 \Lambda_\pi$ in the notation of 
Ref.{\cite {dhr}}.)
This leads to gauge and 
matter couplings that are qualitatively similar to those of the SM 
Higgs boson. The radion mass ($m_r$) is expected to be significantly 
below the scale $\Lambda$ implying that the radion may be 
the lightest new field predicted by the RS model. One may expect on general 
grounds that this mass should lie in the range of  
a few $\times 10$ GeV $\leq m_r \leq \Lambda$. 
The phenomenology of the RS radion has been examined by a number of 
authors{\cite {big}} and in particular has been reviewed for these proceedings 
by Kribs{\cite {Kribs}}. 

\begin{figure}[htbp]
\centerline{
\includegraphics[width=7cm,angle=90]{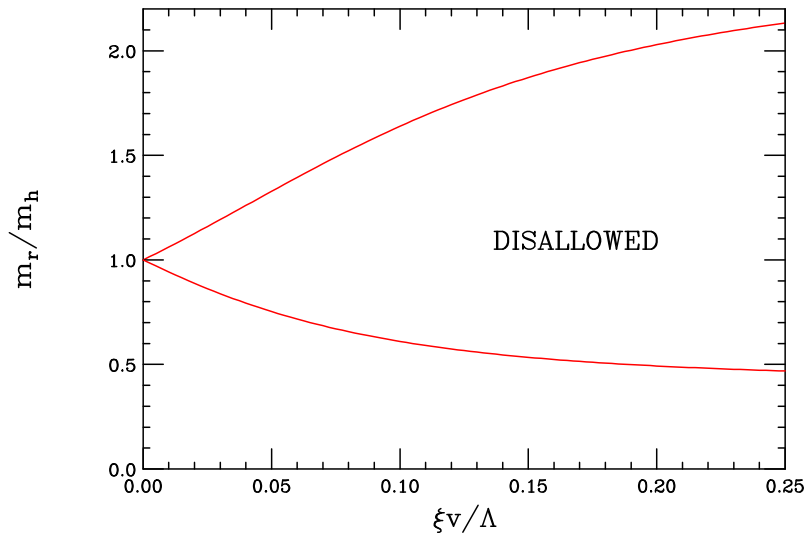}}
\vspace*{0.1cm}
\caption{Constraint on the ratio of the mass of the radion to that of 
the Higgs boson as a 
function of the product $\xi v/\Lambda$ as described in the text. The 
disallowed region lies between the curves.}
\label{p3-38_fig1}
\end{figure}

On general grounds of covariance, the radion may mix with the SM Higgs field on 
the TeV brane through an interaction term of the form 
\begin{equation}
S_{rH}=-\xi \int d^4x \sqrt{-g_w} R^{(4)}[g_w] H^\dagger H\,,
\end{equation}
where $H$ is the Higgs doublet field, 
$R^{(4)}[g_w]$ is the Ricci scalar constructed out of the induced metric $g_w$
on the SM brane,  and $\xi$ is a mixing parameter assumed to be 
of order unity and with unknown sign. The 
above action induces kinetic mixing between the `weak eigenstate' $r_0$ and 
$h_0$ fields which can be removed through a set of field redefinitions and 
rotations. Clearly, since the radion and Higgs boson 
couplings to other SM fields 
differ this mixing will induce modifications in the usual SM expectations for 
the Higgs decay widths. To make unique predictions in this scenario we need to 
specify four parameters: the masses of the {\it physical} Higgs and radion 
fields, $m_{h,r}$, the mixing 
parameter $\xi$ and the ratio $v/\Lambda$, where $v$ is 
the vacuum expectation value of the SM Higgs $\simeq 246$ GeV. Clearly the 
ratio $v/\Lambda$ cannot be too large as $\Lambda_\pi$ is already bounded 
from below by collider and electroweak precision 
data{\cite {dhr}}; for definiteness we will take $v/\Lambda \leq 0.2$ and 
$-1 \leq \xi \leq 1$ in what follows although larger absolute values of $\xi$ 
have been entertained in the literature.  
The values of the two physical masses themselves are not arbitrary. 
When we require that the weak basis mass-squared parameters of the radion and 
Higgs fields be real, as is required by hermiticity, we obtain an additional 
constraint on the ratio of the 
physical radion and Higgs masses which only depends on the product 
$|\xi| {v\over {\Lambda}}$. Explicitly one finds that 
either ${m_r^2\over {m_h^2}}\geq 
1+2\sin^2 \rho+2|\sin \rho| \sqrt{1+\sin^2 \rho}$ or ${m_r^2\over {m_h^2}}\leq 
1+2\sin^2 \rho-2|\sin \rho| \sqrt{1+\sin^2 \rho}$ where 
$\rho=\tan^{-1}(6\xi {v\over {\Lambda}})$. This disfavors the radion having a 
mass too close to that of the Higgs when there is significant mixing; the 
resulting excluded region is shown in Fig.~\ref{p3-38_fig1}. These 
constraints are somewhat restrictive; if we take $m_h=115$ GeV and 
$\xi {v\over {\Lambda}}=0.1(0.2)$ we find that either $m_r>189(234)$ GeV or 
$m_r<70(56)$ GeV. This lower mass range may be disfavored by direct 
searches. 

Following the notation of Giudice \etal {\cite {big}}, the coupling of the 
physical Higgs to the SM fermions and massive gauge bosons $V=W,Z$ 
is now given by
\begin{equation}
{\cal {L}}={-1\over {v}}(m_f\bar ff-m_V^2 V_\mu V^\mu)[\cos \rho \cos \theta +
{v\over {\Lambda}}(\sin \theta-\sin \rho \cos \theta)]h\,,
\end{equation}
where the angle $\rho$ is given above and $\theta$ can be calculated in 
terms of the 
parameters $\xi$ and $v/\Lambda$ and the physical Higgs and radion masses. 
Denoting the combinations $\alpha=\cos \rho \cos \theta$ and 
$\beta=\sin \theta-\sin \rho \cos \theta$, the corresponding Higgs 
coupling to gluons 
can be written as $c_g {\alpha_s\over {8\pi}}G_{\mu\nu}G^{\mu\nu}h$ with 
$c_g={-1\over {2v}}[(\alpha +{v\over {\Lambda}}\beta)F_g
-2b_3\beta {v\over {\Lambda}}]$ where $b_3=7$ is the $SU(3)$ $\beta$-function 
and $F_g$ is a well-known kinematic function of the ratio of masses of the  
top quark to the physical Higgs. 
Similarly the physical Higgs couplings to two photons is now given by 
$c_\gamma {\alpha_{em}\over {8\pi}}F_{\mu\nu}F^{\mu\nu}h$ where 
$c_\gamma={1\over {v}}[(b_2+b_Y)\beta {v\over {\Lambda}}-(\alpha 
+{v\over {\Lambda}}\beta)F_\gamma]$, where $b_2=19/6$ and $b_Y=-41/6$ are the 
$SU(2)\times U(1)$ $\beta$-functions and $F_\gamma$ is another well-known 
kinematic function of the ratios of the $W$ and top masses to the physical 
Higgs mass. (Note that in the simultaneous limits $\alpha \to 1,~\beta \to 0$ 
we recover the usual SM results.) From these expressions we can now compute  
the change of the various decay widths and branching fractions of the
SM Higgs due to mixing with the radion.

\begin{figure}[htbp]
\centerline{
\includegraphics[width=5.4cm,angle=90]{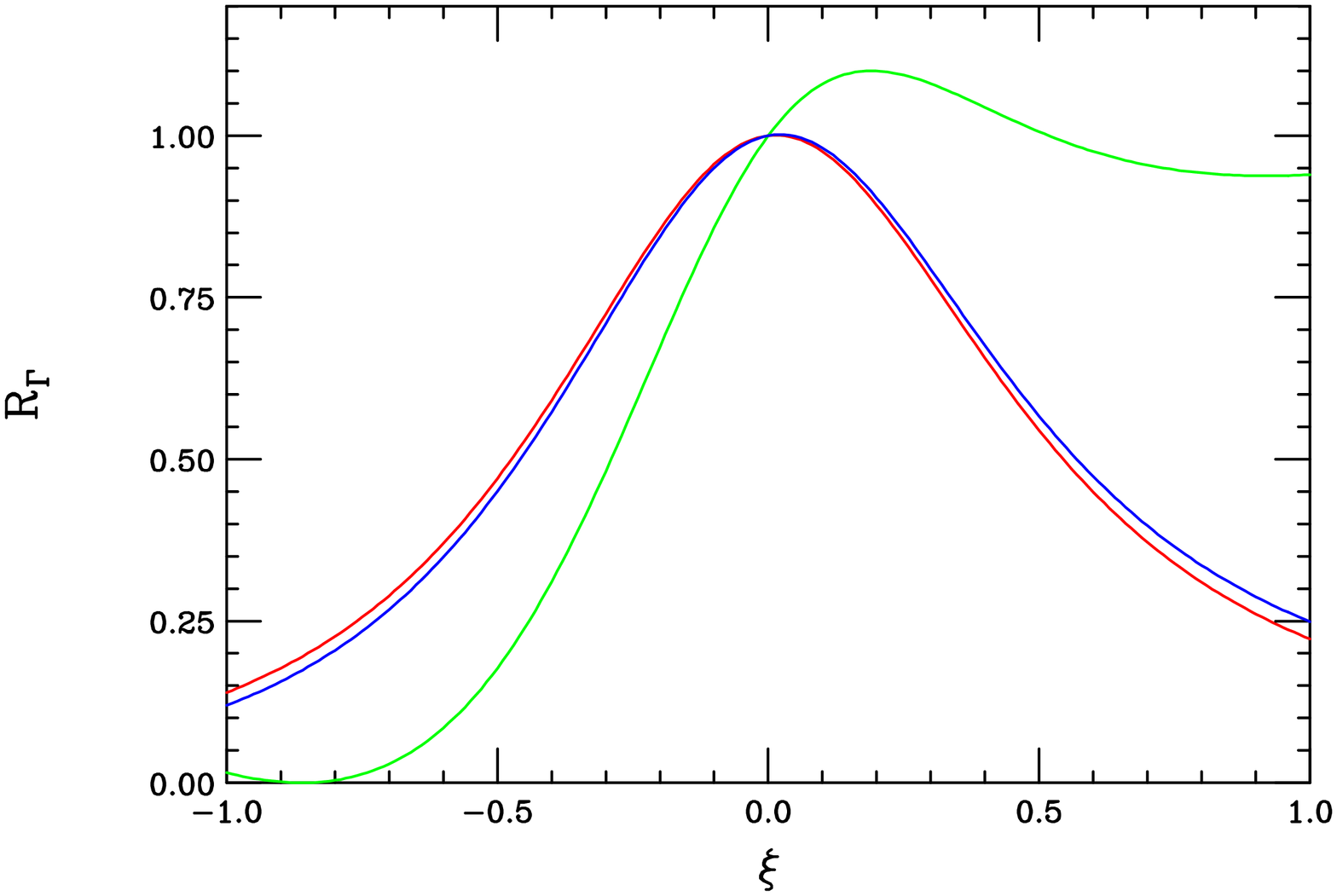}
\hspace*{5mm}
\includegraphics[width=5.4cm,angle=90]{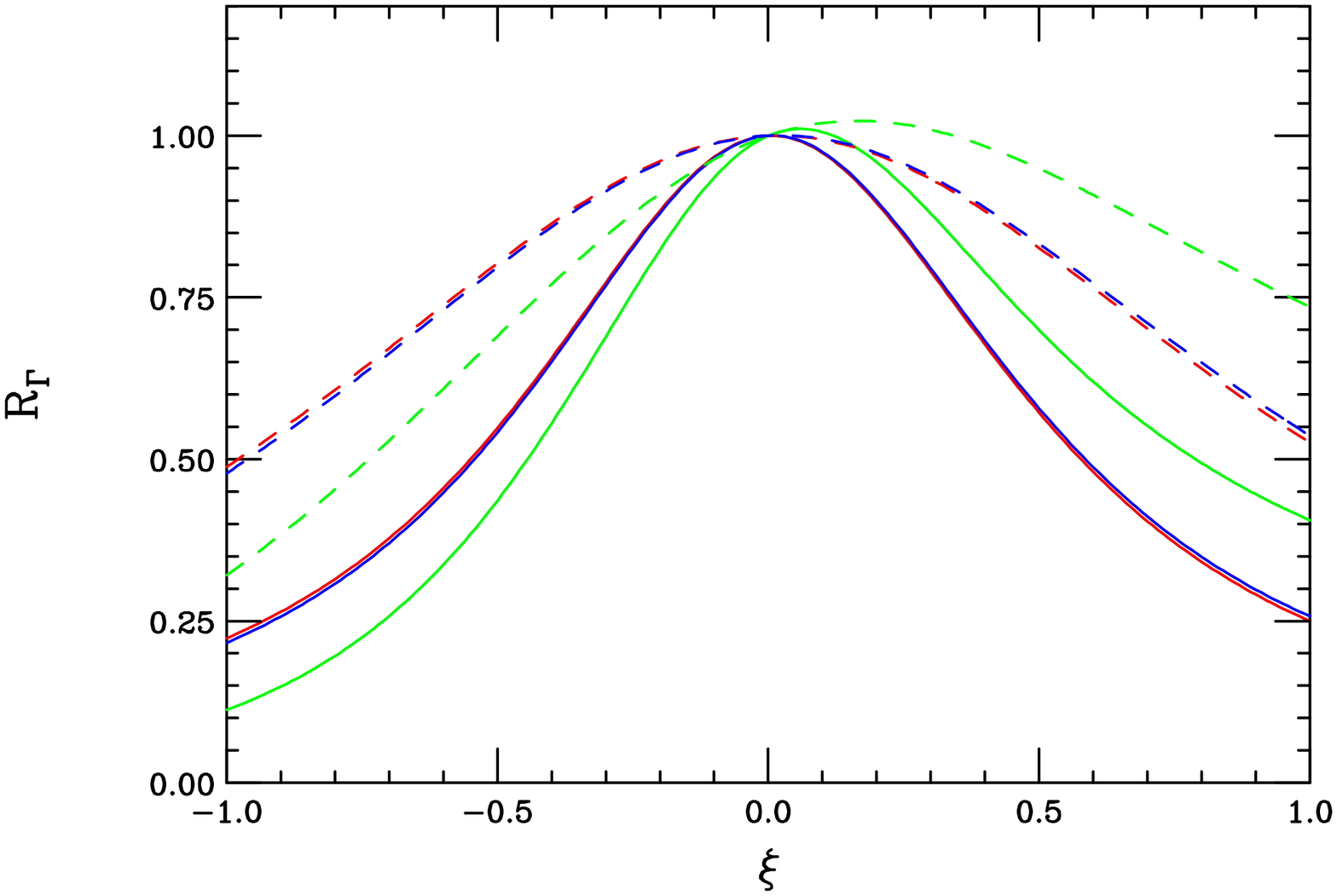}}
\vspace*{0.1cm}
\caption{Ratio of Higgs widths to their SM values, $R_\Gamma$, as a function 
of $\xi$ assuming a physical Higgs mass of 125 GeV: red for fermion pairs or 
massive gauge boson pairs, green for gluons and blue for photons. In the left 
panel we assume $m_r=300$ GeV and $v/\Lambda=0.2$. In the right panel the 
solid(dashed) curves are for $m_r=500(300)$ GeV and $v/\Lambda=0.2(0.1)$.}
\label{p3-38_fig2}
\end{figure}

Fig.~\ref{p3-38_fig2} shows the ratio of the various Higgs widths in 
comparison to their SM expectations as functions of the parameter $\xi$ 
assuming that $m_h=125$ GeV with different values of $m_r$ and 
${v\over {\Lambda}}$. We see several features right away: ($i$) the shifts in 
the widths to $\bar ff/VV$ and $\gamma \gamma$ final states are very similar; 
this is due to the relatively large magnitude of $F_\gamma$ while the 
combination $b_2+b_Y$ is rather small. ($ii$) On the otherhand the shift for 
the $gg$ final state is quite different since $F_g$ is smaller than $F_\gamma$ 
and $b_3$ is quite large. ($iii$) For relatively light radions with a low 
value of $\Lambda$ the width into the $gg$ final state can come close to 
vanishing due to a strong 
destructive interference between the two contributions to the 
amplitude for values of $\xi$ near -1. ($iv$) Increasing the value of $m_r$ 
has less of an effect on the width shifts than does a decrease in the ratio 
${v\over {\Lambda}}$. 

\begin{figure}[htbp]
\centerline{
\includegraphics[width=5.4cm,angle=90]{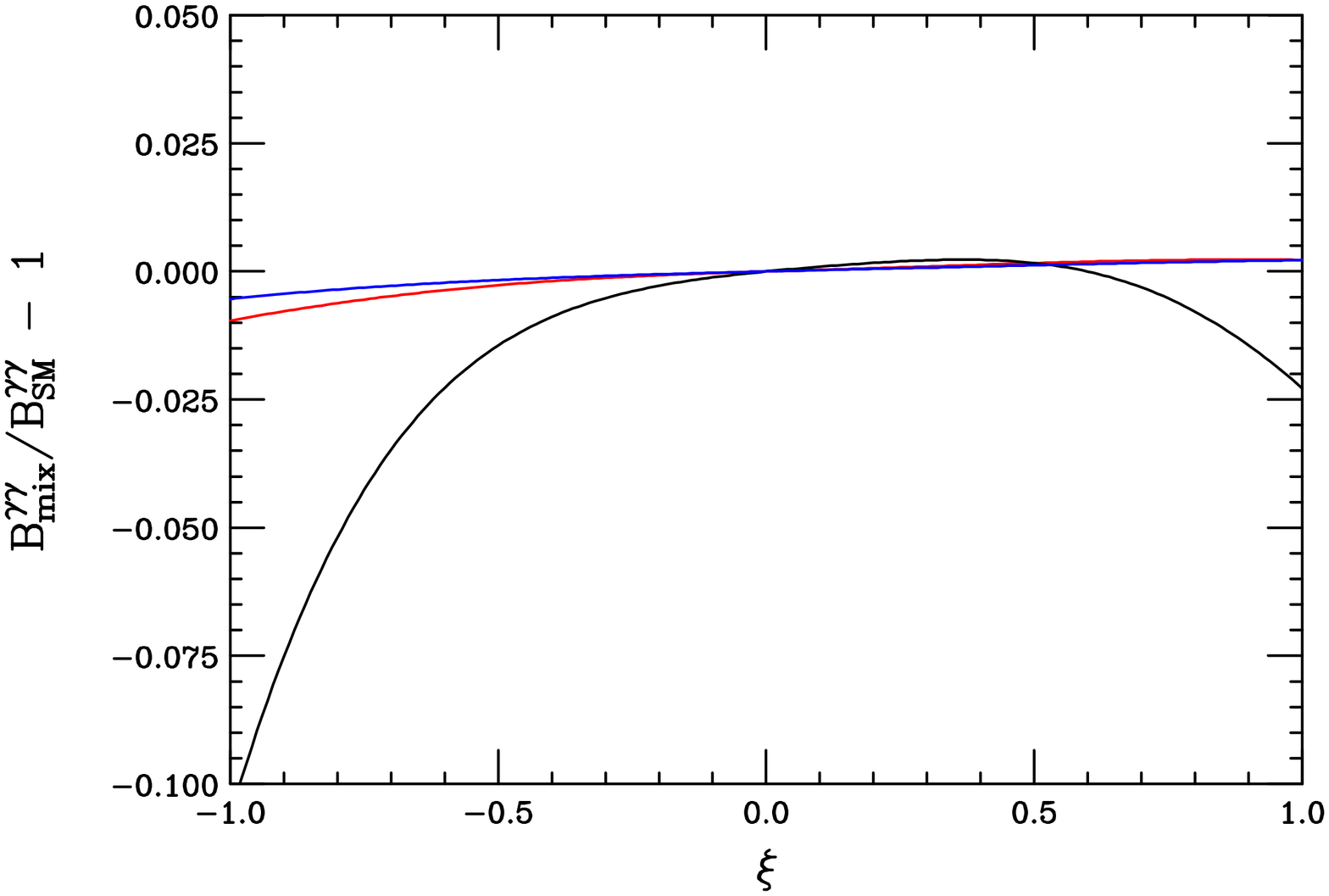}
\hspace*{5mm}
\includegraphics[width=5.4cm,angle=90]{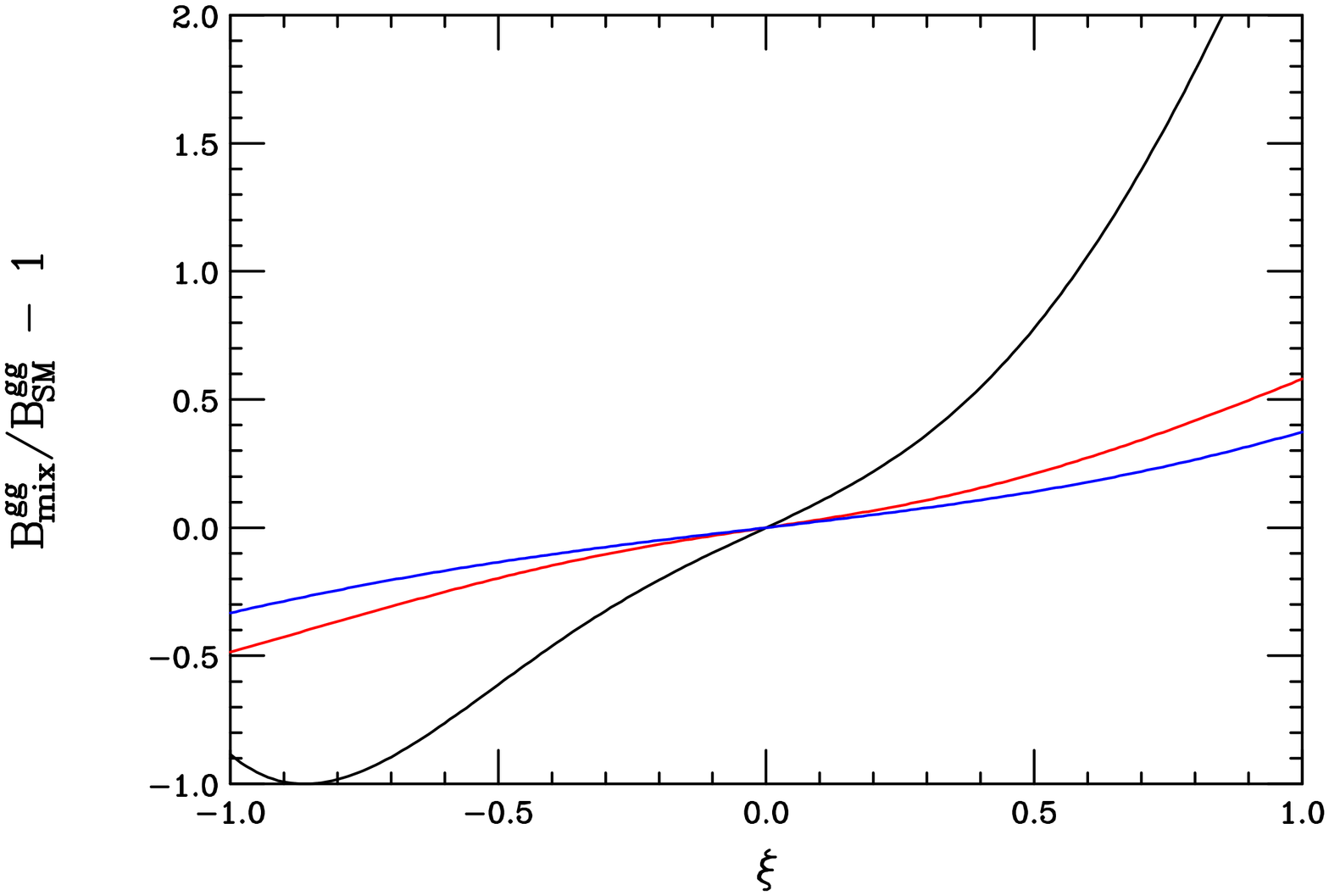}}
\vspace*{0.1cm}
\centerline{
\includegraphics[width=5.4cm,angle=90]{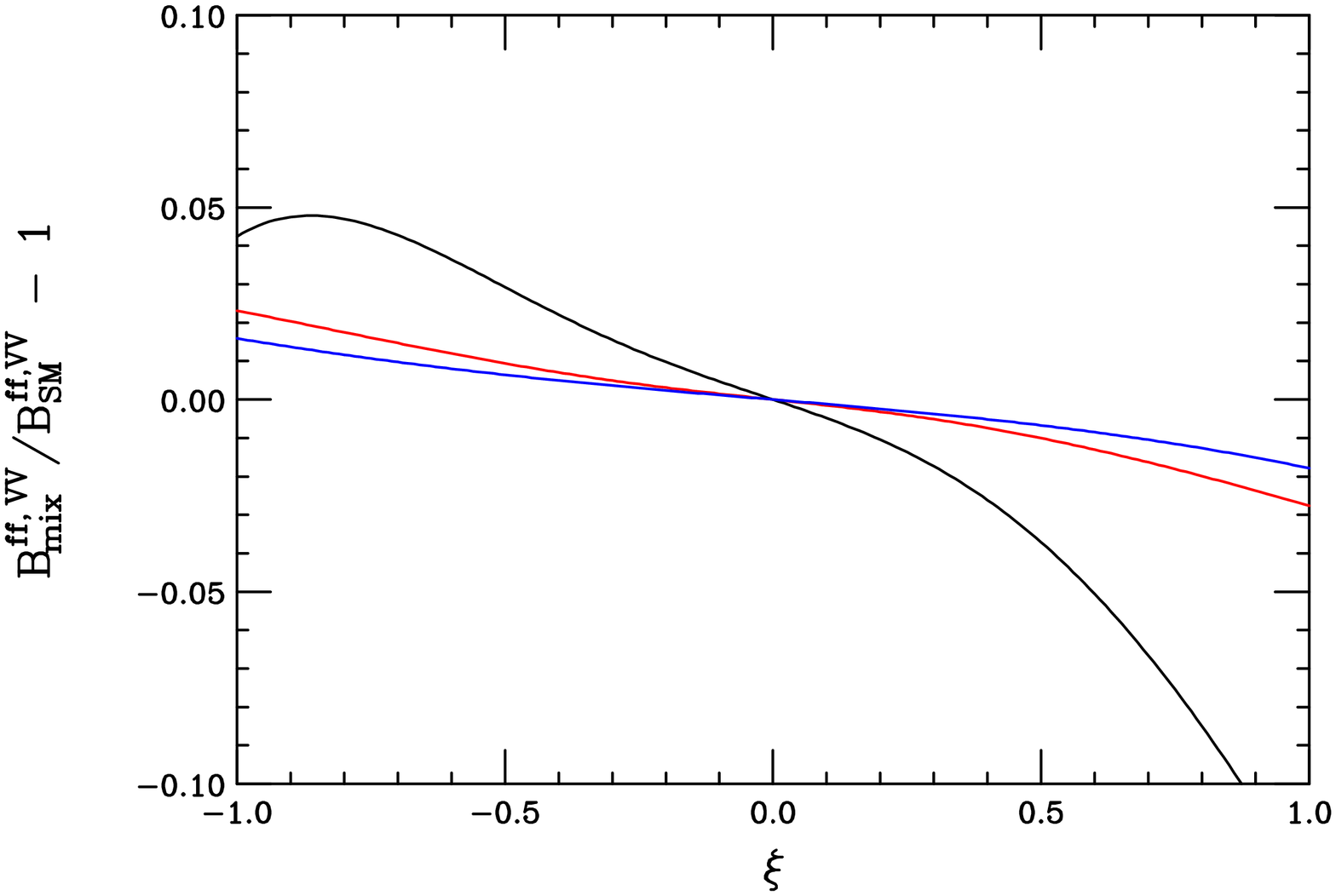}
\hspace*{5mm}
\includegraphics[width=5.4cm,angle=90]{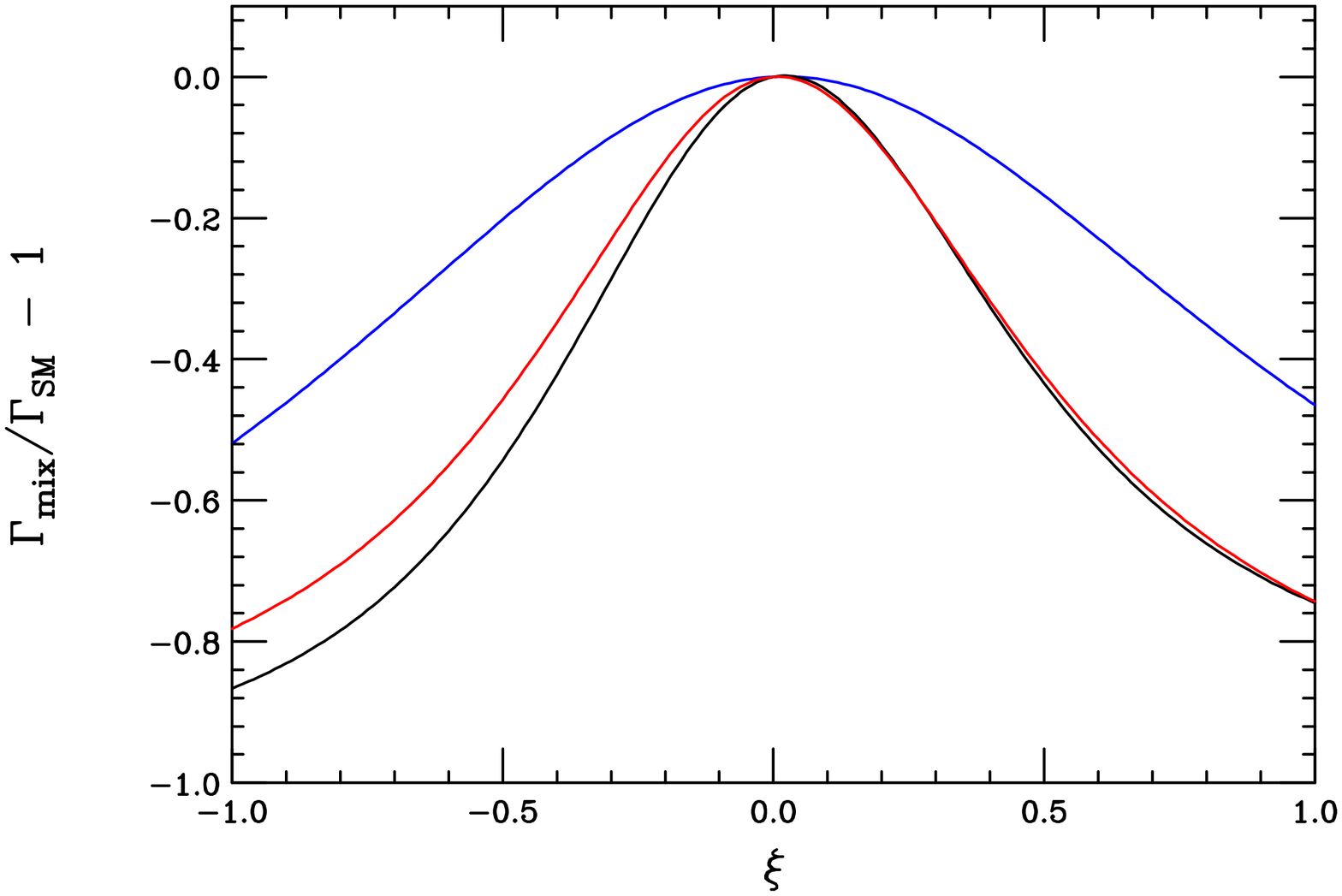}}
\vspace*{0.1cm}
\caption{The deviation from the SM expectations for the Higgs
branching fraction into $\gamma\gamma$, $gg$, $f\bar f$, and $VV$ final
states as labeled, as well as for the total width.  The black, red, and
blue curves correspond to the parameter choices $m_r=300, 500, 300$ GeV
with $v/\xi=0.2, 0.2, 0.1$, respectively.}
\label{p3-38_fig3}
\end{figure}

The  deviation from the SM expectations for
the various branching fractions, as well as the total width, of the 
Higgs are displayed in Fig. \ref{p3-38_fig3} as a function of the 
mixing parameter $\xi$.  We see that the gluon branching fraction and the
total width may be drastically different than that of the SM.  The former
will affect the Higgs production cross section at the LHC.  However, the
$\gamma\gamma$, $f\bar f$, and $VV$, where $V=W,Z$ branching fractions
receive small corrections to their SM values, of order $\lsim 5\%$
for most of the parameter region. Observation of these shifts 
will require the accurate determination
of the Higgs branching fractions available at an $e^+e^-$ Linear Collider.

In summary, we see that Higgs-radion mixing, which is present in some
extra dimensional scenarios, can have a substantial effect on the
properties of the Higgs boson.  These modifications affect the widths
and branching fractions of Higgs decay into various final states, which
in turn can alter the
Higgs production cross section at the LHC and may require the precision
of a Linear Collider to detect.

%
%%%%%%%%%%%%%%%%%%--- References
%%%%%%%%%%%%%%%%%%%%%%%%%%%%%%%%%%%%%%%%%%%%%%%%%%%%%%%
\def\MPL #1 #2 #3 {Mod. Phys. Lett. {\bf#1},\ #2 (#3)}
\def\NPB #1 #2 #3 {Nucl. Phys. {\bf#1},\ #2 (#3)}
\def\PLB #1 #2 #3 {Phys. Lett. {\bf#1},\ #2 (#3)}
\def\PR #1 #2 #3 {Phys. Rep. {\bf#1},\ #2 (#3)}
\def\PRD #1 #2 #3 {Phys. Rev. {\bf#1},\ #2 (#3)}
\def\PRL #1 #2 #3 {Phys. Rev. Lett. {\bf#1},\ #2 (#3)}
\def\RMP #1 #2 #3 {Rev. Mod. Phys. {\bf#1},\ #2 (#3)}
\def\NIM #1 #2 #3 {Nuc. Inst. Meth. {\bf#1},\ #2 (#3)}
\def\ZPC #1 #2 #3 {Z. Phys. {\bf#1},\ #2 (#3)}
\def\EJPC #1 #2 #3 {E. Phys. J. {\bf#1},\ #2 (#3)}
\def\IJMP #1 #2 #3 {Int. J. Mod. Phys. {\bf#1},\ #2 (#3)}
\def\JHEP #1 #2 #3 {J. High En. Phys. {\bf#1},\ #2 (#3)}

\end{document}